\documentclass[review,10pt,1p]{article}

\usepackage{arxiv}

\usepackage[utf8]{inputenc} % allow utf-8 input
\usepackage[T1]{fontenc}    % use 8-bit T1 fonts
\usepackage{amsfonts}       % blackboard math symbols
\usepackage{nicefrac}       % compact symbols for 1/2, etc.
\usepackage{microtype}      % microtypography
\usepackage{booktabs}
\usepackage{lineno,hyperref}
\hypersetup{colorlinks, citecolor=blue, linkcolor=red, urlcolor=Black}
\usepackage[flushleft]{threeparttable}
\usepackage{amsmath,amssymb,amsfonts,bm}
\usepackage{tikz-cd}
\usepackage{graphics}
\usepackage{graphicx}
\usepackage{multirow}
\usepackage{caption}
\usepackage{subcaption}
\usepackage{hyperref}
\usepackage{adjustbox}
\usepackage{mathtools}
\usepackage{booktabs}
\usepackage{tabularx}
\usepackage{multirow}

\usepackage{threeparttable}
\usepackage{geometry}
\usepackage{stmaryrd} % for better symbol variety
\newcommand{\filledstar}{\mathbin{\bigstar}}

\usepackage[linesnumbered,ruled,vlined]{algorithm2e}

\errorcontextlines\maxdimen
\usepackage{amsthm}

\makeatother

\newtheorem{remark}{Remark}[section]

\title{Neural Hodge Corrective Solvers: A Hybrid Iterative–Neural Framework } % Force line breaks with 

\author{
 Arjun Puthli \\
Department Of Applied Mechanics, IIT Delhi \\
  Department of Mathematics\\
  Department of Computer Science and Information Systems\\
    BITS Pilani KK Birla Goa Campus\\
    Sancoale, 403726, India\\
  \texttt{f20212249@goa.bits-pilani.ac.in} \\
   \And
 Somdatta Goswami \\
  Department of Civil and Systems Engineering\\
  Johns Hopkins University\\
  Baltimore, MD, 21218 \\
  \texttt{somdatta@jhu.edu} \\
  \And
    Souvik Chakraborty \\
    Department of Applied Mechanics\\
  Yardi School of Artificial Intelligence (ScAI)\\
  Indian Institute of Technology (IIT) Delhi\\
  Hauz Khas, 110016, India \\
  \texttt{souvik@am.iitd.ac.in}
}

\begin{document}
\maketitle

\begin{abstract}
We introduce the Neural Hodge Corrective Solver (NHCS), a hybrid iterative–neural framework for partial differential equations that embeds learned corrective operators within the Discrete Exterior Calculus (DEC) formulation. The method combines classical Jacobi–Richardson iterations with data-driven corrections to refine numerical solutions while preserving the underlying topological and metric structure.
NHCS employs a two-phase training strategy. In the first phase, DEC operators are learned through relative residual minimization from data. In the second phase, these operators are integrated into the iterative solver, and training targets the improvement of convergence through learned corrective updates that remain effective even for inaccurate intermediate solutions. This staggered training enables stable, progressive refinement while maintaining the structure-preserving properties of DEC discretizations.
To improve multiscale adaptivity, NHCS introduces a convolutional neural network–based correction term capable of capturing fine-scale solution features via localized updates informed by global context, improving scalability over mesh component-wise neural approaches. Moreover, the proposed framework substantially reduces computational cost by avoiding Newton–Raphson–based training and the associated Jacobian evaluations of parameterized operators. The resulting solver achieves improved efficiency, robustness, and accuracy without compromising numerical stability.
\end{abstract}


% \end{frontmatter}

\section{Introduction}
Partial differential equations (PDEs) form the mathematical foundation of a wide range of physical and engineering systems, governing phenomena such as fluid dynamics \cite{Batchelor_2000}, heat transfer \cite{heat-trans}, magnetostatics \cite{Griffiths:1492149,maxwellfem}, and wave propagation \cite{strauss2007partial}. The accurate and efficient solution of PDEs therefore remains a central objective in scientific computing. Classical numerical methods, including finite difference \cite{strikwerdafinite}, finite element \cite{hughes2012finite}, finite volume \cite{fvm}, and spectral methods \cite{canuto2007spectral}, provide well-established and theoretically grounded frameworks with strong guarantees on consistency, stability, and convergence. However, attaining high accuracy with these methods typically requires fine spatial discretizations and, for time-dependent problems, small time steps, resulting in substantial computational cost. This challenge becomes particularly acute in many-query settings, where PDEs must be solved repeatedly for varying initial conditions, boundary conditions, or system parameters, rendering conventional solvers impractical for real-time decision-making or large-scale parametric studies. Motivated by these limitations, data-driven approaches based on deep learning have recently gained attention as a means of amortizing the cost of repeated PDE solves. In particular, neural operator frameworks \cite{neuralop} aim to learn mappings between infinite-dimensional function spaces, thereby approximating the solution operator itself rather than individual solution instances. Early examples such as the Deep Operator Network (DeepONet) \cite{lu_learning_2021}, with its branch–trunk architecture, and subsequent integral-kernel–based models including Fourier \cite{li_fourier_2020}, Wavelet \cite{TRIPURA2023115783}, and Laplace \cite{cao_laplace_2024} neural operators, have demonstrated strong empirical performance as surrogate models for complex physical systems, including heat conduction and fracture mechanics. Despite their success, these approaches largely function as black-box approximators, offering limited interpretability, structure preservation, and theoretical guarantees compared to classical numerical discretizations, thereby highlighting the need for hybrid methodologies that combine the efficiency of learning-based models with the robustness and structure of traditional numerical solvers.

Hybrid methods attempt to close this gap by combining the stability and interpretability of classical numerical methods with the adaptability and computational efficiency of deep learning. Broadly, these methods fall into two categories, namely the Deep Learning-guided (DL-G) methods and the Deep Learning-assisted (DL-A) methods. In DL-G, the neural architectures are employed to adaptively enhance the convergence of iterative solvers. For example, the \textit{flexible conjugate gradient neural operator} \cite{rudikov_fcgno} employs a spectral neural operator \cite{fanaskov_spectral_2023} to predict conjugate gradient update directions, thereby accelerating convergence. The Fourier neural solver \cite{cui_fourier_2022} couples a stationary iterative method with a frequency-space correction module based on Fast Fourier Transform, where the frequency-space filter is predicted by a neural network based on the parameters of the PDE that are taken as inputs. \cite{wang_mg} proposed learning the prolongation operator of the algebraic multigird method by solving a locally-weighted least square model, and applied it to random diffusion equations, and one-dimensional Helmholtz equations with small wave numbers. An additional neural network term acts as a non-linear correction factor for the operator, further accelerating convergence. The deep multigrid method\cite{KATRUTS_dmg} models the convergence of the geometric multigrid method, and then optimizes a deep neural network to mimic a similar convergence pattern. On the other hand, DL-A integrate neural approximators with classical solvers in a more direct manner. Prior to the development of operator learning methods, Int-Deep \cite{huang_intdeep} used a trained deep residual network \cite{resnet} to provide an initial guess that is passed to a conventional finite element solver. The Hybrid Iterative Neural Transferable Solver (HINTS) \cite{kopanicakova_deeponet_2025} alternates between classical updates (e.g., Jacobi or Gauss–Seidel) and a pre-trained DeepONet. HINTS was trained and evaluated on one- and two-dimensional Poisson and Helmholtz equation datasets, with regular and irregular geometries. \cite{hu_hybrid_2025} extended this approach by replacing the DeepONet with a MIONet \cite{jin_mionet_2022}, enabling the handling of multiple input functions. These hybrid approaches aim to achieve a balance between rigorous convergence guarantees and the adaptability of learned models. Although they have been shown to generalize across different mesh discretizations, the deep learning components still function as black-box approximators.

Discrete Exterior Calculus (DEC) \cite{hirani_thesis, desbrun_discrete_2005}  provides a coordinate-free, geometrically consistent framework for representing differential operators on discrete meshes, and has found extensive success in computational graphics applications \cite{crane_geom, desburn_geom, Mohamed04052018}. A distinguishing feature of DEC is that it naturally encodes vector calculus conservation principles through discrete analogues of fundamental theorems such as Stokes’ theorem. This makes DEC particularly appealing for designing PDE solvers, as it enforces essential physical conservation properties at the discrete level. Consequently, DEC-based solvers preserve geometric and topological structures of the underlying domain, ensuring that numerical approximations remain faithful to physical laws \cite{arnold_finite_2018}. These methods have been extensively applied in the literature for various applications in fluid flow problems \cite{WANG2023112245, MOHAMED_NSDEC}. \cite{TRASK2022110969} introduced a data-driven DEC framework for solving PDEs, where metric information is learned from data and combined with purely topological DEC operators. While their approach demonstrated the promise of data-guided geometric learning through numerical experiments on the two-dimensional Darcy flow and Magnetostatics problems, it primarily served as a proof of concept and exhibited notable scalability limitations.

In response to this challenge, we present the \textbf{Neural Hodge Corrective Solver (NHCS)}, a DL-assisted hybrid method that incorporates learned corrective operators within the DEC framework, and updates solution estimates via a combination of Richardson and Jacobi iterations. Our key contributions in this work are primarily algorithmic in nature and are as follows:
\begin{itemize}
\item \textbf{Two Phase Training:} Our staggered two phase training approach involves first learning the DEC operators by relative residual minimization using the given data. Next, the learned operators are used within a hybrid Jacobi–Richardson  framework to progressively refine the solution obtained using the numerical solver. Training in this phase enables the model to improve its predictions iteratively, even when the current estimate is inaccurate, ensuring robust convergence while preserving the underlying topological and metric structure.
\item \textbf{Improved Multiscale Adaptivity:} A CNN-based correction term is introduced to improve the model's capacity to capture fine-scale variation in the solution space, in contrast to the mesh component-wise neural networks used in \cite{TRASK2022110969}. This unified convolutional architecture enables the correction term to learn localized adjustments informed by global context, while improving scalability and efficiency.
\item \textbf{Reduced Computational Cost:} The training method employed in \cite{TRASK2022110969} incurs high computational cost by simultaneously performing Newton–Raphson iterations and updating model parameters via gradient descent, requiring repeated Jacobian (or Jacobian vector product) computations of the parameterized differential operators for each Newton-Raphson update. Our two-phase training strategy, combined with Richardson-like iterative updates, eliminates the need for these expensive Jacobian evaluations, significantly reducing the overall computational cost while maintaining accuracy and stability.
\end{itemize}

The remainder of the paper is structured as follows: In Section~\ref{sec:proposed}, we introduce the NHCS method, and review essential background material. Section~\ref{sec:num} presents the evaluation pipeline and numerical experiments to validate the efficacy and robustness of our method. The problems considered are multiscale Poisson and Helmholtz equations, and a linear elasticity problem on a 2D domain with geometrical discontinuity. Finally, Section~\ref{sec:conclusion} reviews the salient features and key findings of our method. We conclude by going over potential future research directions.

\section{Neural Hodge Corrective Solver}\label{sec:proposed}
\subsection{Classical DEC Formulation}
\noindent Consider a PDE of the form: 
\begin{equation}
\begin{alignedat}{2}
\mathcal{L}(\bm u) & = f \quad && \text{in }\Omega, \\
g(\bm u) &= \bm{u_b} \quad && \text{on }\partial\Omega,
\end{alignedat}
\end{equation}
where $f$ is a forcing function term, $u_b$ is some boundary condition, $\bm u$ is the solution we are interested in, and $\mathcal{L}$ is a linear elliptic differential operator. To discretize $\mathcal{L}$ in a structure-preserving manner, we employ Discrete Exterior Calculus (DEC).

Let \(\Omega\) denote a simplicial complex representing the computational domain. A \(k\)-chain \(C_k\) is a linear combination of oriented \(k\)-simplices, and the \textbf{boundary operator} \(\partial_k: C_{k+1} \to C_k\) maps each \((k+1)\)-simplex to the oriented sum of its \(k\)-dimensional faces. The dual space of \(k\)-chains is the space of \(k\)-cochains \(C^k: C_k \to \mathbb{R}\). The \textbf{coboundary operator} \(\delta_k: C^k \to C^{k+1}\) encodes how local quantities accumulate along higher-dimensional simplices. Boundary and coboundary operators satisfy a discrete analogue of Stokes’ theorem,
\begin{equation}
\langle \delta_k \alpha, \beta \rangle = \langle \alpha, \partial_k \beta \rangle, \quad \alpha \in C^k, \ \beta \in C_{k+1},
\end{equation}
and satisfy \(\delta_{k+1} \circ \delta_k = 0\), corresponding to classical vector calculus identities (e.g., \(\mathrm{curl} \circ \mathrm{grad} = 0\) for \(k=0\)).
The \textbf{discrete Hodge star} operator \(\filledstar_k: C^k \to C^{n-k}\) introduces the metric structure:
\begin{equation}
(\filledstar_k)_{ii} = \frac{|D(\sigma_i^k)|}{|\sigma_i^k|},
\end{equation}
where \(D(\sigma_i^k)\) is the dual element corresponding to \(\sigma_i^k\) and \(|\cdot|\) denotes its geometric measure. It defines a discrete inner product
\begin{equation}
\langle x, y \rangle_{\filledstar_k} = x^\top \filledstar_k y,
\end{equation}
and induces the \textbf{codifferential} \(\delta_k^* = \filledstar_k^{-1} \delta_k^\top \filledstar_{k+1}\). Combining these yields the \textbf{discrete Hodge Laplacian}:
\begin{equation}
L_k = \delta_{k-1} \delta_{k-1}^* + \delta_k^* \delta_k = \delta_k \filledstar_{k-1}^{-1} \delta_{k-1}^\top \filledstar_k + \filledstar_k^{-1} \delta_k^\top \filledstar_{k+1} \delta_k.
\end{equation}
While DEC preserves topological and metric structure, classical discretizations may converge slowly and fail to capture fine-scale variations in multiscale problems.
\subsection{Data-Driven DEC Operators (DDEC)}

To overcome these limitations, we introduce \textbf{data-driven counterparts} of the DEC operators. The \textbf{data-driven coboundary} and \textbf{codifferential} are defined as
\begin{equation}\label{datacobound}
d_k := B_{k+1} \delta_k B_k^{-1}, \quad d_k^* := D_k^{-1} \delta_k^\top D_{k+1},
\end{equation}
where \(B_k\) and \(D_k\) are diagonal matrices with positive entries (enforced via squaring trainable parameters). The \textbf{data-driven inner product} is
\begin{equation}
\langle x, y \rangle_{D_k B_k^{-1}} := \langle D_k B_k^{-1} x, y \rangle,
\end{equation}
which preserves the adjoint relation between \(d_k\) and \(d_k^*\) and generalizes \(\langle \cdot, \cdot \rangle_{\filledstar_k}\) from classical DEC.
In \(\mathbb{R}^2\), we have
\begin{equation}
GRAD = B_1 \delta_0 B_0^{-1}, \quad CURL = B_2 \delta_1 B_1^{-1}, \quad GRAD^* = D_0^{-1} \delta_0^\top D_1, \quad CURL^* = D_1^{-1} \delta_1^\top D_2,
\end{equation}
and \(DIV = -GRAD^*\). These operators satisfy
\begin{equation}
d_{k+1} \circ d_k = 0, \quad d_k^* \circ d_{k+1}^* = 0.
\end{equation}

\begin{remark}
The learnable matrices \(\mathbf{B_k}\) and \(\mathbf{D_k}\) act as data-driven Hodge stars, capturing metric information from the primal mesh without explicit dual mesh construction.
\end{remark}

The \textbf{data-driven Hodge Laplacian} is
\begin{equation}
\Delta_k =
\begin{cases}
d_0^* d_0, & k = 0, \\
d_{k-1} d_{k-1}^* + d_k^* d_k, & k \ge 1.
\end{cases}
\end{equation}

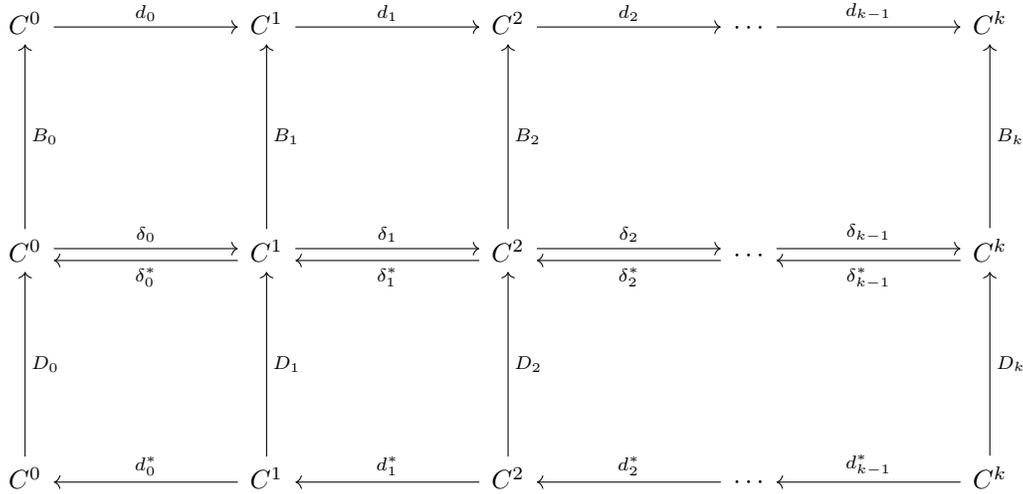
\begin{figure}[htbp]
\centering
\begin{tikzcd}[row sep=7em, column sep=7em]
C^0 \arrow[r, "d_0"] & C^1 \arrow[r, "d_1"] & C^2 \arrow[r, "d_2"] & \cdots \arrow[r, "d_{k-1}"] & C^k \\
C^0 \arrow[r, "\delta_0", shift left=0.5ex] \arrow[r, "\delta_0^*"', leftarrow, shift right=0.5ex] \arrow[u, "B_0"'] 
& C^1 \arrow[r, "\delta_1", shift left=0.5ex] \arrow[r, "\delta_1^*"', leftarrow, shift right=0.5ex] \arrow[u, "B_1"'] 
& C^2 \arrow[r, "\delta_2", shift left=0.5ex] \arrow[r, "\delta_2^*"', leftarrow, shift right=0.5ex] \arrow[u, "B_2"'] 
& \cdots \arrow[r, "\delta_{k-1}", shift left=0.5ex] \arrow[r, "\delta_{k-1}^*"', leftarrow, shift right=0.5ex] 
& C^k \arrow[u, "B_k"'] \\
C^0 \arrow[u, "D_0"'] & C^1 \arrow[l, "d_0^*"'] \arrow[u, "D_1"'] 
& C^2 \arrow[l, "d_1^*"'] \arrow[u, "D_2"'] 
& \cdots \arrow[l, "d_2^*"'] 
& C^k \arrow[l, "d_{k-1}^*"'] \arrow[u, "D_k"']
\end{tikzcd}
\caption{Commutative diagram showing relationships between classical and data-driven DEC operators.}
\label{commdiag}
\end{figure}

Figure~\ref{commdiag} summarizes the relationships between classical and data-driven DEC operators.

\subsection{Neural Hodge Corrective Solver and Training}

To enhance multiscale adaptivity, we augment the DDEC  operator representation  with a CNN-based corrective term:
\begin{equation}\label{datahodge}
\tilde{\Delta_k} = \Delta_k + \mathcal{CNN}(\Delta_k),
\end{equation}
where the CNN captures localized variations across the mesh. Spectral bias in deep networks favors low-frequency components \cite{Rahaman2018OnTS, XU2025113905}, complementing iterative solvers that rapidly reduce high-frequency errors \cite{HARIMI}. 
The CNN consists of four 2D convolutional layers (3×3 kernels, ReLU) with a single-channel output.

\begin{figure}[htbp]
\centering
\includegraphics[width=1\textwidth]{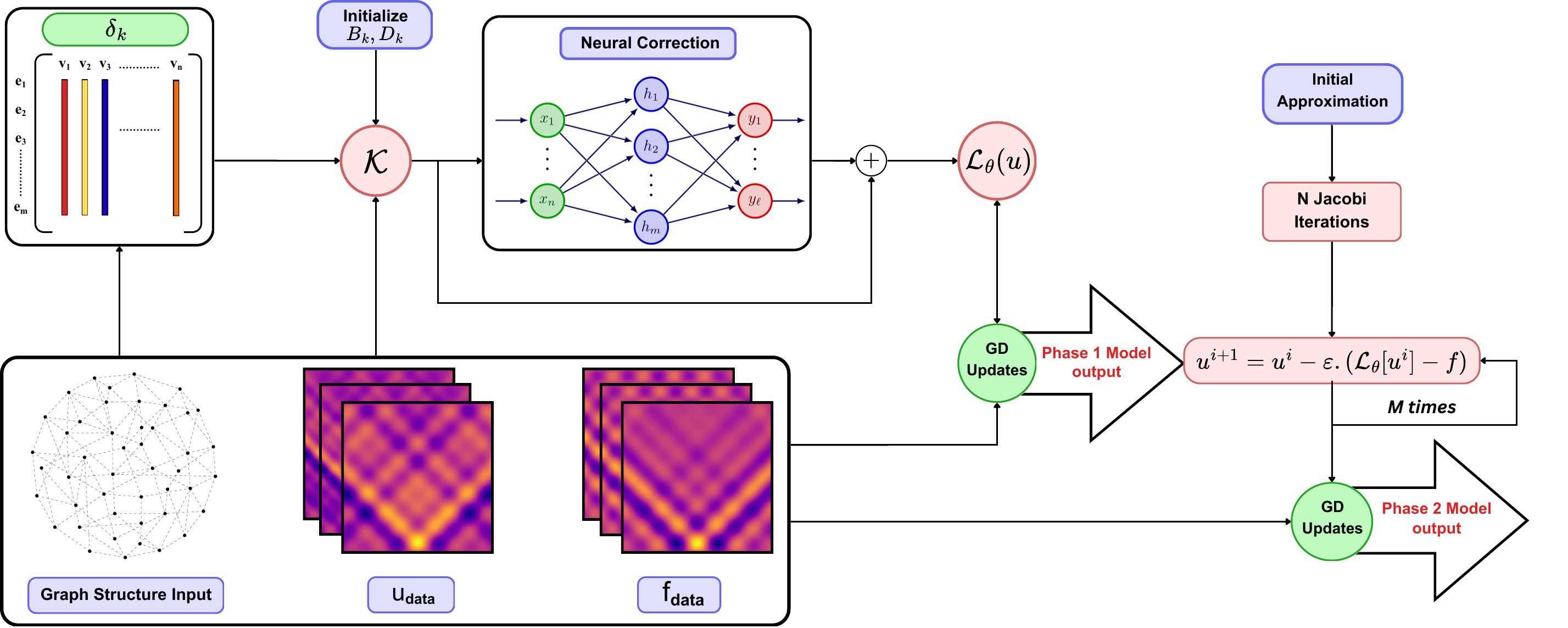}
\caption{NHCS training pipeline for Phase 1 and Phase 2. Phase 1 minimizes \(\| f_{\text{est}} - f_{\text{data}} \|\) to update DDEC operators, while Phase 2 optimizes the hybrid Jacobi + Neural Hodge correction solver.}
\label{fig:eval}
\end{figure}

Figure~\ref{fig:eval} illustrates the training and evaluation pipeline, Algorithm~\ref{algo_train} outlines the training procedure, and Algorithm~\ref{algo2} details the iterative evaluation procedure. We will collectively refer to our model's parameters as $\bm \theta$, which comprises of $\mathbf B_k, \mathbf D_k$ , and the CNN parameters $\boldsymbol \eta$ lumped together. We collectively denote the PDE-specific and training parameters by $\mathcal{P}$. Furthermore, to avoid notational ambiguity, $\bm u^i$ will represent the $i^{th}$ iteration approximation of the solution, and $\bm u_i$ will represent the $i^{th}$ ground-truth solution from the dataset $\mathcal{D}$. In the first training phase, we focus on learning an accurate representation of the operators, by jointly updating the metric representation parameters, $\mathbf {B_k} \ \text{and} \ \mathbf {D_k}$, and the neural correction term. This ensures that the residual, $\mathcal{L}[\bm u] -f$ is small enough, providing a stable operator representation for the next stage. In the second phase, we integrate our operator into an iterative solving framework. Starting with an initial guess for the solution, it is refined through \(\mathcal{N}\) Jacobi iterations, which rapidly reduce high-frequency errors. This is followed by \(\mathcal{M}\) Neural Hodge corrective iterations (i.e. $\mathcal{L}(\bm u^n) - f$), which capture multiscale features and low-frequency components that classical iterations alone cannot resolve. By integrating structure-preserving data-driven DEC operators with a convolutional correction term, the solver achieves rapid convergence with minimal computational overhead. The staggered two-phase training ensures stable and interpretable learning of the data-driven operators, while the evaluation procedure avoids costly Jacobian computations and reduces memory usage. Note that $GD(L, \theta)$ represents the gradient descent step where we are updating parameters $\theta$, and minimizing the loss $L$. The loss function, we employed, is the relative mean square error (Relative MSE), defined as:
\begin{equation}
\mathrm{L}(x,y)
= \frac{\frac{1}{N}\sum_{i=1}^N (x_i - y_i)^2}
       {\frac{1}{N}\sum_{i=1}^N y_i^2}.
\end{equation}
to account for varying scales in the data due to coefficient randomness. Overall, this framework provides a unified, lightweight, and highly adaptive PDE solver capable of efficiently resolving complex multiscale variations without sacrificing mathematical rigor or efficiency.

\begin{algorithm}[!h]
    \LinesNotNumbered
    \caption{NHCS Training}
    \label{algo_train}
    \KwData{PDE and training parameters $\mathcal{P}$, Graph Structure $\mathcal{G}$}
    \tcp{Phase 1:}
    \textbf{Initialization:} $\leftarrow$ Coboundary and Codifferential operators $\delta_k, \ \delta_k^*$, Convolutional Neural Network $\mathcal{CNN}$, model parameters $\theta$ \\
    \For{$e \leftarrow 1$ \KwTo \text{epochs}}{
        $\mathcal{D} \leftarrow GenerateDataset(\mathcal{P})$ \\
        \For{$u_i, f_i \in \mathcal{D}$}{
            $f_{est} = \mathcal{L}(u_i)$ \\
            $\text{loss} \leftarrow L(f_i, \ f_{est})$\\
            $\theta \leftarrow GD(\text{loss}, \theta)$
        }
    } 
    \tcp{Phase 2:}
    \textbf{Initialization:} $\leftarrow$ $\varepsilon$, Jacobi Iterations \textbf{I}, Neural-Hodge Iterations  \textbf{M}, Jacobi stencil \textbf{A} \\
    \For{$e \leftarrow 1$ \KwTo \textit{epochs} }{
        $\mathcal{D} \leftarrow GenerateDataset(\mathcal{P})$ \\
        \For{$u_i, f_i \in \mathcal{D}$}{
            $u^0 \leftarrow 0$ \\

            \tcp{Perform \textbf{I} Jacobi iterations}
            \For{$n \leftarrow 1$ \KwTo \textbf{I}}{
                $u^n \leftarrow \ u^{n-1} + A^{-1}(f_i - A u^{n-1})$ 
            }
            
            \For{$n \leftarrow 1$ \KwTo \textbf{M}}{
                $f^{est} \ = \mathcal{L}(u^{I+n-1})$ \\
                $u^{I+n} \leftarrow u^{I+n-1} \ - \ \varepsilon.(f^{est} - f_i)$
            }
            $\text{loss} \leftarrow L(u^{I+N}, \ u_i)$ \\
            $\theta \leftarrow GD(\text{loss}, \theta)$
        }
    }
    \Return{$\theta$}
\end{algorithm}

\begin{algorithm}[!h]
\LinesNotNumbered
\caption{NHCS Testing}
\label{algo2}
\KwIn{PDE parameters \(\mathcal{P}\), forcing \(f\), Jacobi stencil \(A\), learned NHCS model \(\mathcal{L}\), iteration configuration \((N,M)\), step size \(\varepsilon\)}
\textbf{Initialize:} \(u^0 \leftarrow 0\) \\[2pt]

\tcp{Jacobi pre-iterations}
\For{$i \leftarrow 1$ \KwTo $N$}{
    $u^i \leftarrow u^{i-1} + A^{-1}(f - A u^{i-1})$ 
}

\tcp{Learned correction iterations}
\For{$j \leftarrow 1$ \KwTo $M$}{
    $u^{i+j} \leftarrow u^{i+j-1} - \varepsilon (\mathcal{L}(u^{i+j-1}) - f)$
}

\Return{$u^{N+M}$}
\end{algorithm}

\section{Numerical Experiments}
\label{sec:num}

To validate our proposed method, we conduct numerical experiments on three canonical elliptic PDEs: the Poisson equation, the Helmholtz equation, and the linear elastic equation. These equations are ubiquitous in computational science and engineering, arising in contexts as diverse as incompressible fluid dynamics, electrostatics, acoustics, electromagnetics, and structural design. In incompressible Navier–Stokes solvers, for instance, the pressure Poisson equation and Helmholtz-type momentum equation constitute the dominant computational bottleneck, while in wave propagation and scattering problems, the Helmholtz equation governs frequency-domain behavior. Owing to their prevalence across domains and their analytical tractability, these equations serve as ideal test cases to examine the effectiveness and robustness of our approach. We construct synthetic datasets by prescribing analytical Fourier-type solutions with randomized coefficients and deriving the corresponding forcing terms exactly. This guarantees precise ground-truth supervision while introducing controlled variability to prevent overfitting. Finally, we demonstrate the effectiveness of the proposed framework on a physically motivated problem, namely the two-dimensional linear elasticity equation, thereby highlighting its generality and applicability.

Since we are not applying any external forces in the elasticity problem, we use the root mean square error (RMSE) for Phase 1 training. We have employed the AdamW optimizer (with a weight decay of $1\times10^{-5}$), which decouples weight decay from gradient updates \cite{loshchilov2018decoupled}, instead of Adam, as it provides a more stable learning curve for our model. For the Poisson and Helmholtz equation we also made use of the Reduce LR on Plateau (RoP) scheduler. The training configurations across the three problems are summarised in Table \ref{tab:training_params}. All training and evaluation was done on a single NVIDIA GeForce RTX-3050 GPU with $4GB$ of RAM. 

\begin{table}[h!]
\centering
\renewcommand{\arraystretch}{1.2}
\setlength{\tabcolsep}{6pt}
\begin{tabular}{l|cc|cc|cc}
\hline
\multirow{2}{*}{\textbf{Training Parameter}} 
& \multicolumn{2}{c|}{\textbf{Poisson}} 
& \multicolumn{2}{c|}{\textbf{Helmholtz}} 
& \multicolumn{2}{c}{\textbf{Elasticity}} \\ 
\cline{2-7}
& \textbf{Phase 1} & \textbf{Phase 2} 
& \textbf{Phase 1} & \textbf{Phase 2} 
& \textbf{Phase 1} & \textbf{Phase 2} \\ 
\hline
Epochs            & 120 & 60 & 120 & 30 & 500 & 1000 \\
Optimizer         & AdamW & AdamW & AdamW & AdamW & AdamW & AdamW \\
LR Scheduler      & RoP & RoP & RoP  & RoP & - & - \\
Initial LR        & $1\times10^{-2}$ & $1\times10^{-2}$ & $1\times10^{-2}$ & $1\times10^{-2}$ & $1\times10^{-4}$ & $1\times10^{-4}$ \\
Minimum LR        &  $5\times10^{-6}$ &   $1\times10^{-5}$  &  $5\times10^{-6}$   &   $1\times10^{-5}$  &  -   &  -   \\
Scheduler Reduction Factor      &  $0.8$  &   $0.8$  &  $0.8$   &   $0.8$  &  -   &  -   \\
Scheduler Tolerance     &  $0.1$  &   $0.02$  & $0.1$   &   $0.02$  &  -   &  -   \\
Richardson update step $\varepsilon$     &  -  &  $2.5\times10^{-5}$   &  -    &  $1.5\times10^{-2}$  &  -   &  $1\times10^{-1}$   \\
\hline
\end{tabular}
\caption{Training configurations for the Poisson, Helmholtz, and Elasticity problems across both training phases. RoP is the Reduce LR on Plateau scheduler.}
\label{tab:training_params}
\end{table}

\begin{figure}[!h]
    \centering
    \includegraphics[width=1\textwidth]{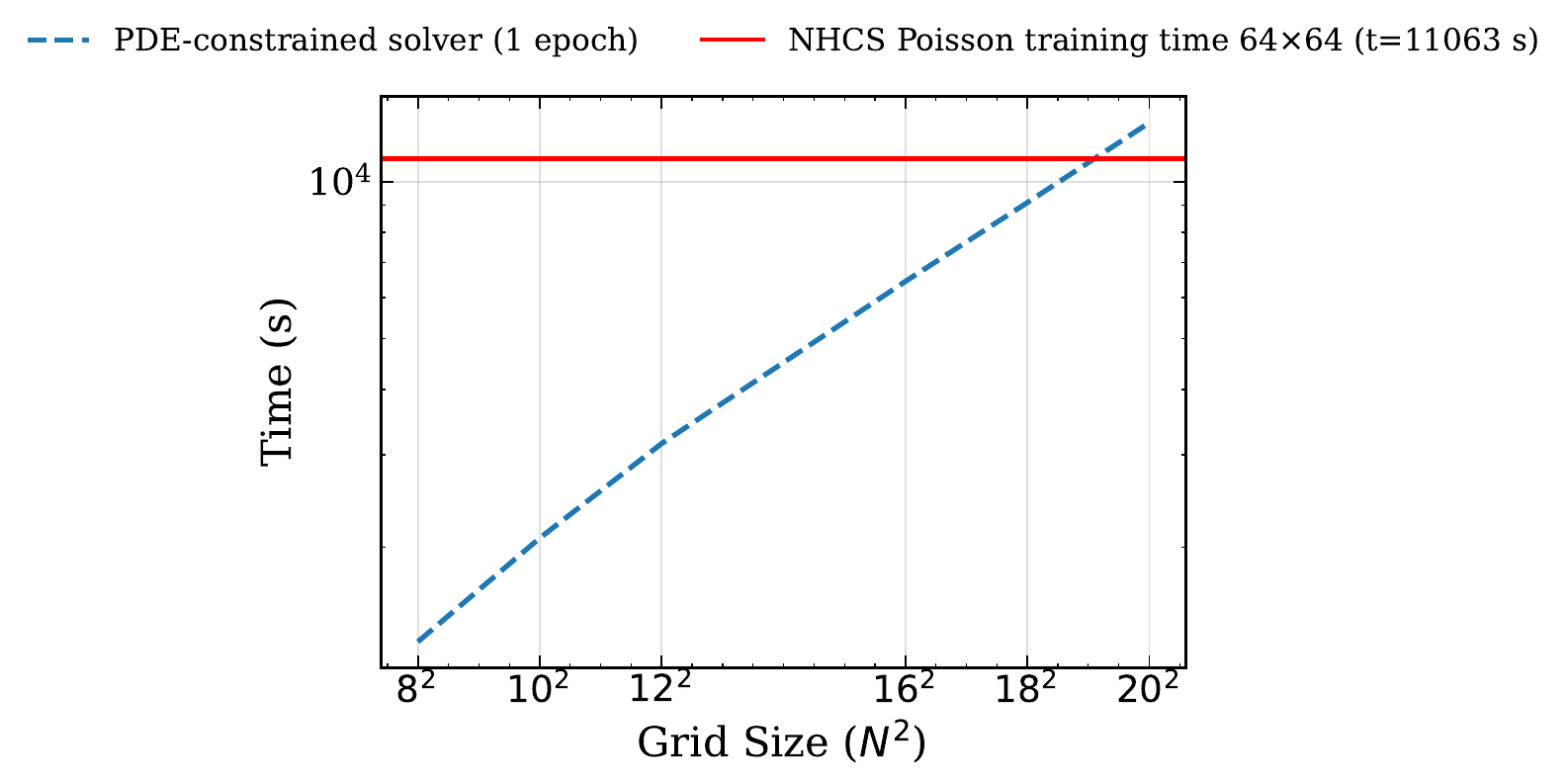}
    \caption{Poisson equation training time plot. Training time (in seconds) for 1 epoch for the PDE constrained optimization algorithm in \cite{TRASK2022110969} with varying grid size, represented by the blue dashed line. The component wise neural networks have one hidden linear layer, with a width of $10$, and ReLU activation functions between successive layers. A dataset with $512$ samples, and batch size $1$ is used. The red line represents $t=11063s$, the time taken for Phase 1 + Phase 2 of training for the $64\times64$ model.} 

    \label{fig:pdeopt}
\end{figure}

Figure \ref{fig:pdeopt} shows the time taken per epoch for the PDE optimization algorithm presented in \cite{TRASK2022110969} on the Poisson equation dataset of size 512. The perturbation networks used in this scenario have one hidden layer, with a width of 10, and ReLU activations between successive layers. This highlights the substantial computational cost of the original DDEC algorithm, showing that the entire training procedure of NHCS requires significantly less time ($4480$ sec) than even a single training epoch of the original DDEC method. For a fair comparison, a dataset of the same size is used for NHCS training as well.

We compare our proposed hybrid model with the HINTS algorithm~\cite{hybriddeeponet}. The DeepONet employed within the HINTS framework consists of two subnetworks: a \emph{Branch Net} and a \emph{Trunk Net}. The Branch Net consists of a series of convolutional layers with channel sizes $[2,\,40,\,60,\,100,\,180]$, a kernel size of $3$, and a stride of $2$. The resulting feature map is flattened and passed through three fully connected layers with ReLU activations, producing an output of size $80$. The Trunk Net consists of three fully connected layers of size $80$ with Tanh activations between layers.  Training was carried out for $10{,}000$ epochs using the Adam optimizer with an initial learning rate of $1\times10^{-3}$. A step-based learning-rate scheduler was used, reducing the learning rate by a factor of $0.5$ every $100$ epochs, with a minimum learning-rate bound of $1\times 10^{-5}$. The same architecture is used for the Elasticity problem, except the input dimensions of the Branch Net also accommodates the Young's modulus, Poisson's ratio and top layer displacement. The DeepONet was trained for $20,000$ epochs with a learning rate of $5\times10^{-3}$. These architecture details along with training hyperparameters for the DeepONet are summarized in tabular form in \ref{sec:app1}. HINTS was evaluated with parameters recommended in \cite{hybriddeeponet}, using fixed correction frequency (the first and every 200-th iteration thereafter) to avoid favoring NHCS.

We compare the two methods using the final relative mean squared error (Relative MSE) of the computed solutions and by examining the evolution of the relative residuals throughout the solver process. The Relative MSE is defined as
\begin{equation}
\mathrm{Relative\; MSE}(\bm u_{\text{pred}}, \bm u_{\text{ref}})
= \frac{\frac{1}{N}\sum_{i=1}^{N}\left(u_{\text{pred},i}-u_{\text{ref},i}\right)^2}
{\frac{1}{N}\sum_{i=1}^{N}u_{\text{ref},i}^2},
\end{equation}
where N is the number of spatial grid points that we evaluate for. For a linear system \(\bf A u = f\), the relative residual at iteration \(k\) is defined as
\begin{equation}
r_k = \frac{\|\bf A u^{(k)} - f\|_2}{\|\bf f\|_2}.
\end{equation}

\subsection{Poisson Equation}

\noindent The 2D Poisson equation is written as: 
\begin{equation}\label{poisson}
\begin{alignedat}{2}
\Delta u(x,y) & = -f(x,y) \quad && \forall \;\; (x,y)\in \Omega, \\
u(x,y)        & = 0  \quad && \forall \;\; (x,y)\in \partial\Omega,
\end{alignedat}
\end{equation}
where $u(x,y)$ is a scalar function defined on a 2D bounded domain, and $f(x,y)$ is the forcing function of the PDE. We synthetically generate the dataset using the following analytical representation:
\begin{equation}\label{fourier-poisson}
\begin{aligned}
u(x,y) &= \frac{1}{N}\sum_{n=1}^N a_n \sin(n\pi x)\cos(n\pi y), \\
f(x,y) &= \frac{1}{N}\sum_{n=1}^N 2a_n (n\pi)^2 \sin(n\pi x)\cos(n\pi y),
\end{aligned}
\end{equation}
where $x \in [0,1]$, $y \in [-0.5, 0.5]$, and $N=15$. This adheres to the prescribed form in Equation \ref{poisson}. To generate random $(u,f)$ pairs for the training data, we sample each $a_n$ uniformly from $(0,1)$. Our PDE representation for the above, in data-driven DEC terms, is quite straightforward, and is effectively Equation \ref{datahodge} with $k=0$. The discrete Hodge Laplacian equation is written as:
\begin{equation}
    \mathcal{L}(u) \ = \ \Delta_0u \ + \ \mathcal{CNN}(\Delta_0u).
\end{equation}
The corresponding Jacobi iteration stencil with grid spacing $h$ reads:
\vspace{0.3cm}
\begin{equation}
\frac{1}{h^2}
\begin{bmatrix}
0 & -1 & 0 \\
-1 & 4 & -1 \\
0 & -1 & 0
\end{bmatrix}.
\end{equation}

\begin{figure}[!ht]
    \centering
    \includegraphics[width=1\textwidth]{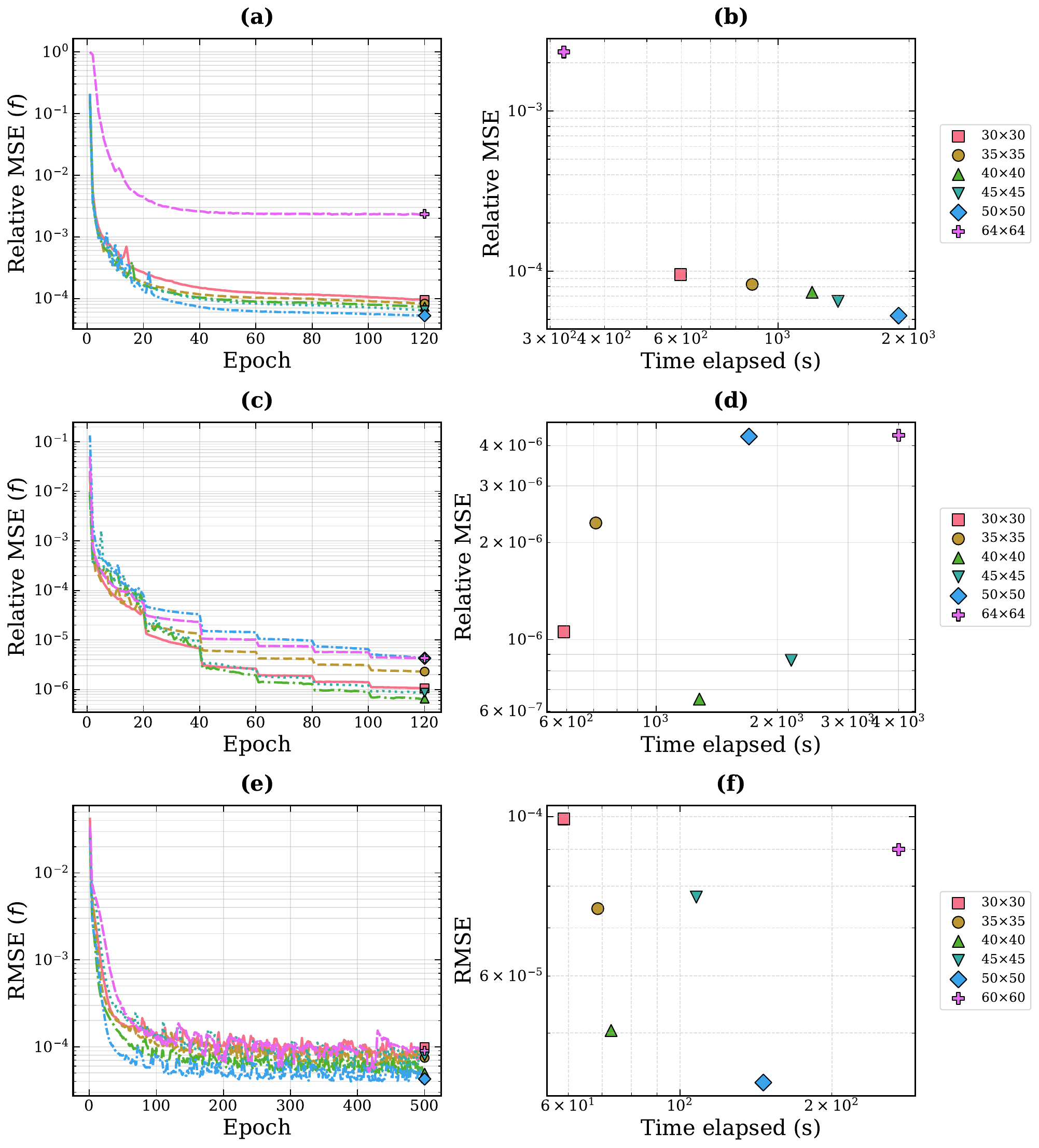}
    \caption{Phase 1 Training performance on the 2D Poisson Equation (a, b), 2D Helmholtz equation (c, d), and the  2D Linear Elasticity Problem (e,f).  (a, c) Relative MSE of the forcing term vs. epochs. (b, d) Relative MSE of the forcing term vs. training time (in seconds). For the Helmholtz problem, we train for wave numbers $\in \{4, 8, 16, 20, 24, 28\}$. (e) RMSE of the forcing term vs. epochs. (f) RMSE of forcing term vs. training time (in seconds).}

    \label{fig:train}
\end{figure} 

\begin{table}[!ht]
    \centering
    \begin{tabularx}{\textwidth}{c *{4}{>{\centering\arraybackslash}X}}
        \toprule
        \multirow{2}{*}{Grid Size} & \multicolumn{2}{c}{Poisson} & \multicolumn{2}{c}{Helmholtz} \\
        \cmidrule(lr){2-3} \cmidrule(lr){4-5}
         & Relative MSE & Time (s) & Relative MSE & Time (s) \\
        \midrule
        $30 \times 30$ & $9.498 \times 10^{-5}$ & $597.65$  & $1.059 \times 10^{-6}$ & $590.25$ \\
        $35 \times 35$ & $8.271 \times 10^{-5}$ & $872.72$  & $2.303 \times 10^{-6}$ & $708.58$ \\
        $40 \times 40$ & $7.379 \times 10^{-5}$ & $1197.85$ & $6.538 \times 10^{-7}$ & $1281.26$ \\
        $45 \times 45$ & $6.470 \times 10^{-5}$ & $1373.65$ & $8.630 \times 10^{-7}$ & $2166.07$ \\
        $50 \times 50$ & $5.273 \times 10^{-5}$ & $1892.27$ & $5.493 \times 10^{-6}$ & $1644.41$ \\
        $64 \times 64$ & $2.339 \times 10^{-3}$ & $4045.40$ & $4.309 \times 10^{-6}$ & $4001.35$ \\
        \bottomrule
    \end{tabularx}
    \caption{Phase 1 training performance (Relative MSE and training time) for Poisson and Helmholtz equations across different grid sizes over 120 epochs.}
    \label{tab:training_comparison}
\end{table}

\begin{figure}[!ht]
    \centering
    \includegraphics[width=1\textwidth]{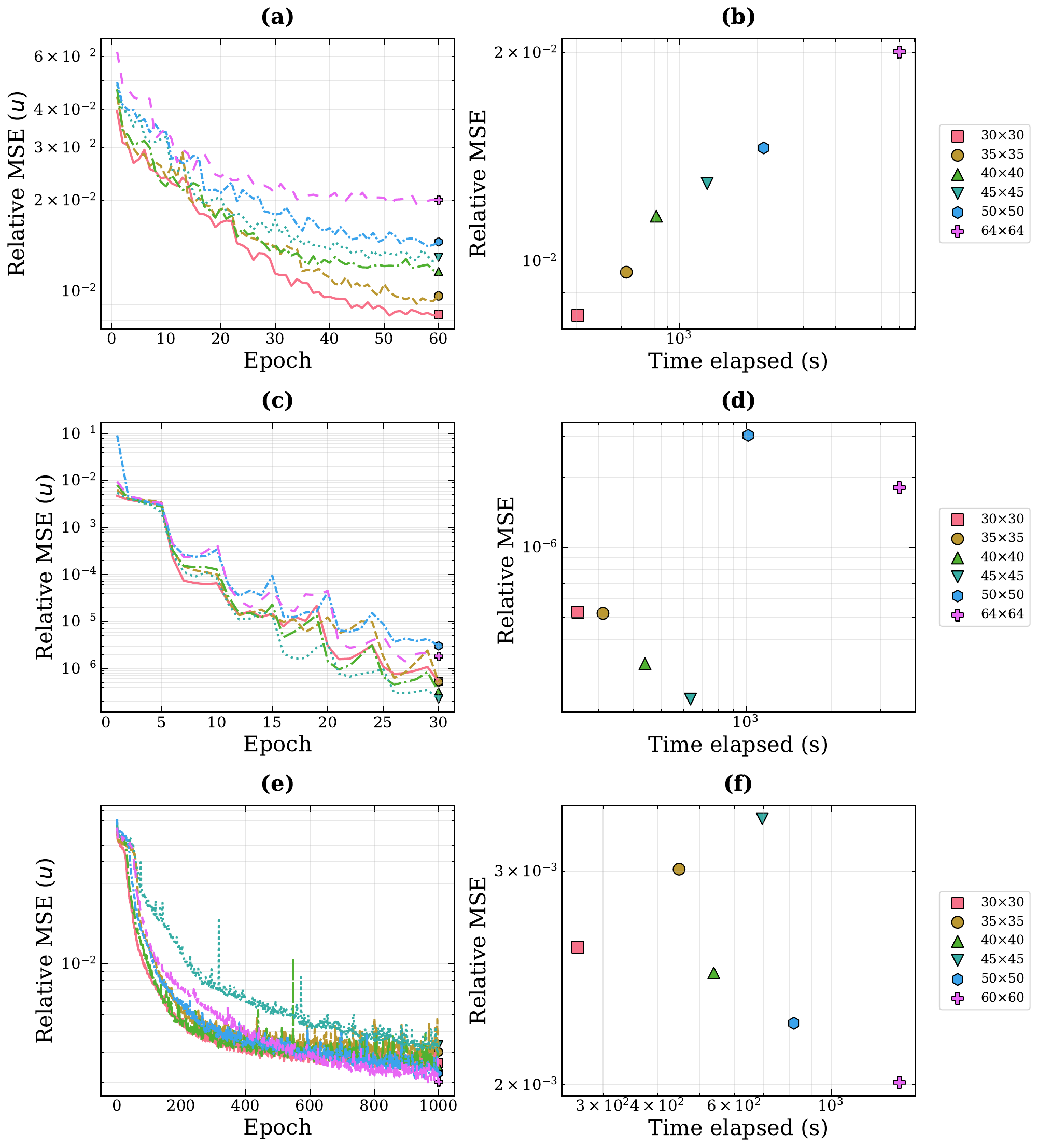}
    \caption{(Phase 2 Training performance on the 2D Poisson Equation (a, b), 2D Helmholtz equation (c, d), and the  2D Linear Elasticity Problem (e,f).  (a, c, e) Relative MSE of the solution vs. epochs. (b, d, f) Relative MSE of the solution vs. training time (in seconds).}

    \label{fig:train2}
\end{figure}

\begin{table}[!ht]
    \centering
    \begin{tabularx}{\textwidth}{c *{4}{>{\centering\arraybackslash}X}}
        \toprule
        \multirow{2}{*}{Grid Size} & \multicolumn{2}{c}{Poisson} & \multicolumn{2}{c}{Helmholtz}\\
        \cmidrule(lr){2-3} \cmidrule(lr){4-5}
         & Relative MSE & Time (s) & Relative MSE & Time (s) \\
        \midrule
        $30 \times 30$ & $8.341 \times 10^{-3}$ & $407.00$  & $5.282 \times 10^{-7}$ & $253.82$ \\
        $35 \times 35$ & $9.639 \times 10^{-3}$ & $625.36$  & $5.208 \times 10^{-7}$ & $311.43$ \\
        $40 \times 40$ & $1.161 \times 10^{-2}$ & $815.10$  & $3.163 \times 10^{-7}$ & $438.67$ \\
        $45 \times 45$ & $1.295 \times 10^{-2}$ & $1279.23$ & $2.230 \times 10^{-7}$ & $635.65$ \\
        $50 \times 50$ & $1.457 \times 10^{-2}$ & $2111.71$ & $3.028 \times 10^{-6}$ & $1017.62$ \\
        $64 \times 64$ & $2.006 \times 10^{-2}$ & $7017.64$ & $1.805 \times 10^{-6}$ & $3490.56$ \\
        \bottomrule
    \end{tabularx}
    \caption{Phase 2 training performance (Relative MSE and training time) for Poisson and Helmholtz equations across different grid sizes over 60 and 30 epochs respectively.}
    \label{tab:training_comparison2}
\end{table}

\begin{figure}[!h]
    \centering
    \includegraphics[width=1\textwidth]{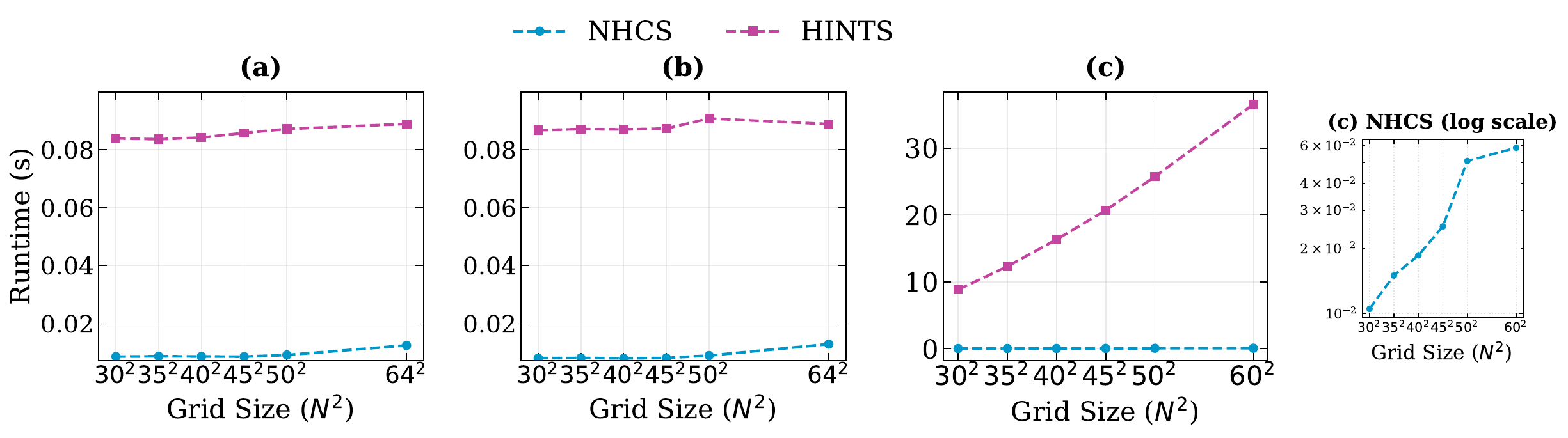}
    \caption{Evaluation runtime comparison between NHCS and HINTS for (a) Poisson Equation, (b) Helmholtz Equation and (c) Elasticity equation. The right-most plot represents the log-scale visualization of the NHCS runtime in (c). Across all three problems, we can observe that NHCS has a faster runtime than HINTS. The difference is especially pronounced in the elasticity problem (c), as the Jacobi solver employed in HINTS has to accommodate the hole boundary in its computation, while the learned DEC operator in NHCS accounts for the domain structure.}
    \label{fig:runtime}
\end{figure}
\begin{figure}[!h]
    \centering
    \includegraphics[width=1\textwidth]{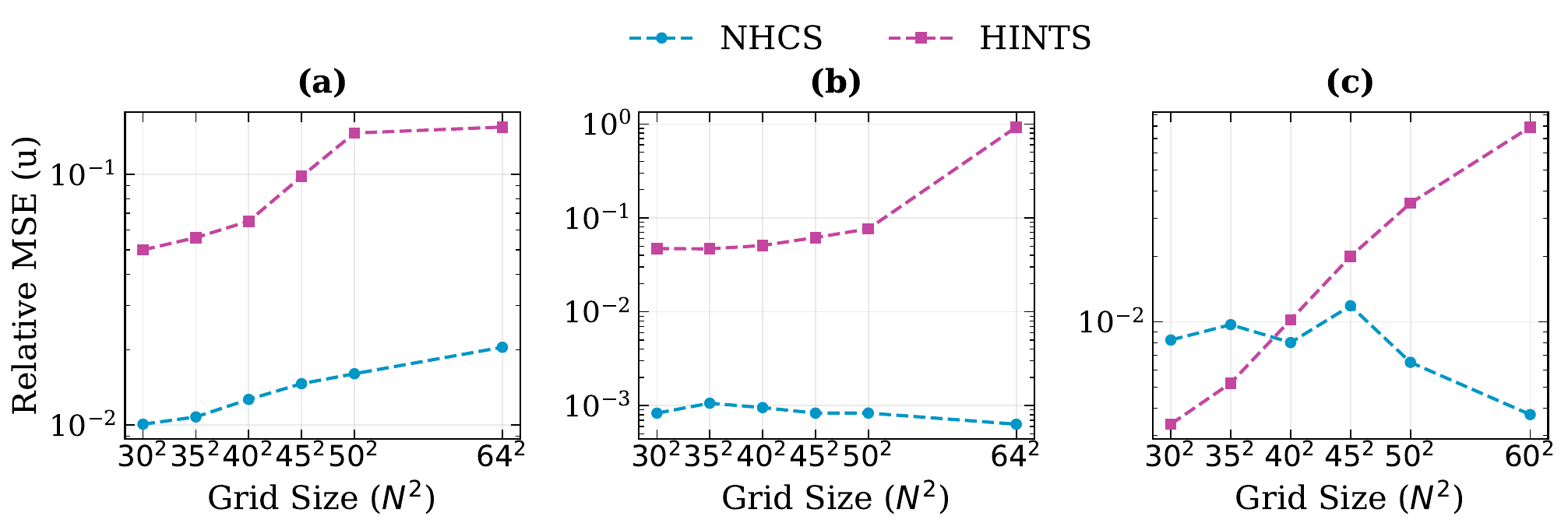}
    \caption{Results Relative MSE comparison between NHCS and HINTS for (a) Poisson Equation, (b) Helmholtz Equation and (c) Elasticity equation. NHCS consistently outperforms HINTS across mesh resolutions for the Poisson (a) and Helmholtz (b) equations. HINTS initially outperforms NHCS in the elasticity equation (c) due to the low frequency nature of the problem, with the Jacobi component of the solver being able to resolve errors better on coarser grids.}
    \label{fig:comparison}
\end{figure}
\begin{figure}[!h]
    \centering
    \includegraphics[width=1\textwidth]{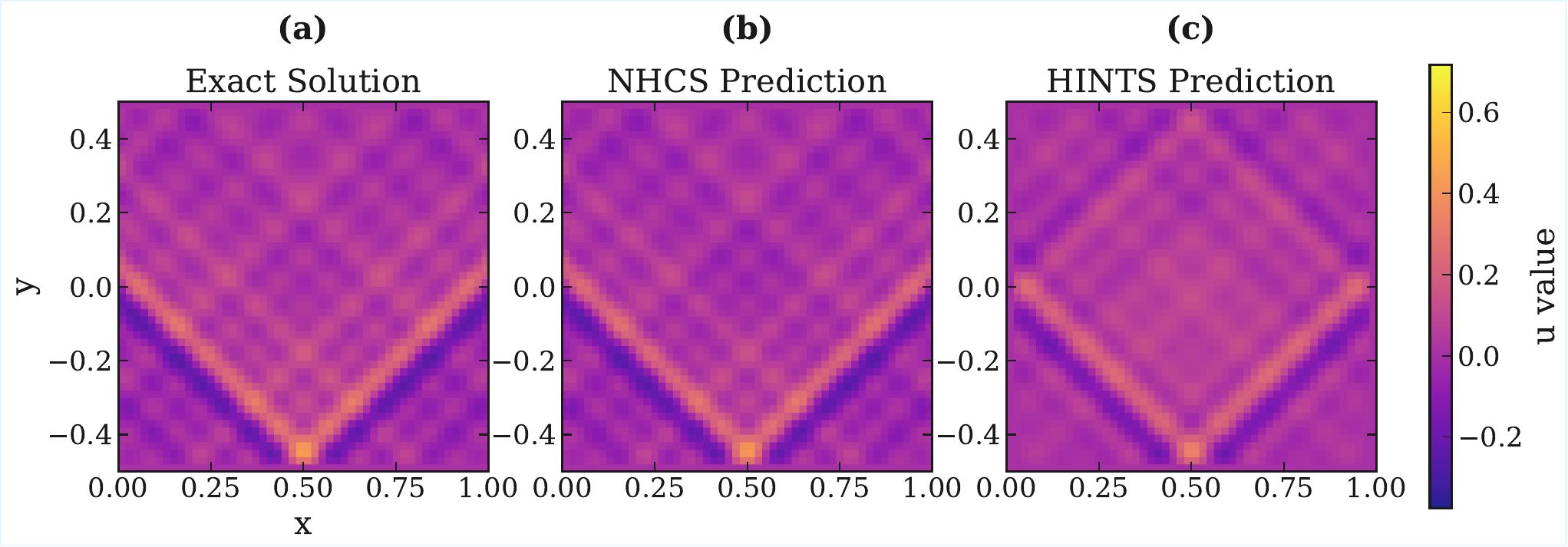}
    \caption{Poisson equation solution estimate comparison between (b) NHCS, and (c) HINTS with respect to the exact solution (a) at a $50\times50$ resolution. The NHCS approximation (b) is much closer to the ground truth (a), as compared to the HINTS approximation (c) which tends to overly smooth predictions. NHCS is able to clearly capture finer variations and local gradients with more precision.}
    \label{fig:poisson_solns}
\end{figure}
\begin{figure}[!h]
    \centering
    \includegraphics[width=1\textwidth]{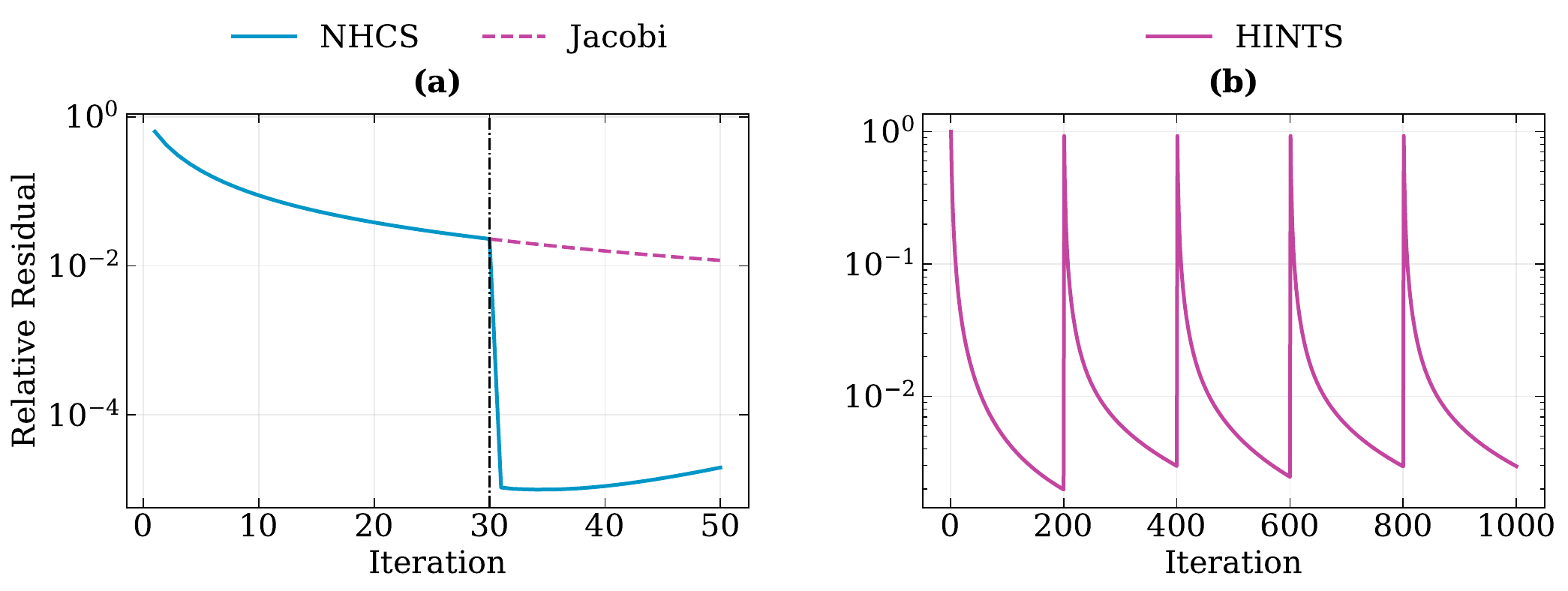}
    \caption{Poisson equation relative residual vs iteration count for NHCS (a) and HINTS (b). In (a) the blue line represents the residual norm of NHCS, the black dotted line represents the iteration where the DEC based Richardson updates take over the Jacobi updates, and the pink dotted line represents the continued residual trajectory of the Jacobi iterations. The sharp fall in the residual in (a) can be attributed to the learned DEC operators contributing to the approximation correction, and the spikes in (b) occur during DeepONet corrections in HINTS.}
    \label{fig:poisson_res}
\end{figure}

Figure~\ref{fig:train}(a,b), and Table~\ref{tab:training_comparison} summarizes the Phase 1 training performance across different mesh sizes. As expected, the time taken for training increases with the discretization density of the mesh. Figure~\ref{fig:train2}(a, b), and Table~\ref{tab:training_comparison2} summarize the Phase 2 training performance, and a $\mathcal{N}:\mathcal{M}$ = $30:20$ configuration is used. Figure~\ref{fig:runtime}(a) illustrates the computational time required to evaluate solutions as a function of grid size. A gradual increase in runtime is observed for HINTS as the grid resolution increases from $30 \times 30$ to $64 \times 64$. In contrast, the runtime of NHCS remains nearly constant over the same range of grid sizes, indicating weak sensitivity to grid refinement. Figure~\ref{fig:comparison}(a) compares the evaluation performance between NHCS and HINTS for the Poisson equation problem. Both methods show a gradual increase in error under mesh refinement, as expected from the final Phase 2 training plots. Specifically, HINTS shows an increase in relative error from approximately $0.05$ at a $30 \times 30$ grid to over $0.1$ at $64 \times 64$, while NHCS observes an increase from just under $0.01$ up to $0.02$. Figure~\ref{fig:poisson_solns} illustrates the performance of the two methods by visualizing the predicted solutions for a randomly generated sample. Both methods capture the overall structure of the solution; however, NHCS exhibits more accurate field approximation, with fewer artifacts and closer agreement to the ground truth compared to HINTS. Moreover, HINTS tends to produce overly smoothed predictions, whereas NHCS, despite not always capturing the potential field precisely, successfully reproduces local gradients. Figure \ref{fig:poisson_res} visualizes the evolution of the residual norm with respect to the iteration count. In \ref{fig:poisson_res}(a) the blue line represents the relative residual of NHCS, the black dotted line represents the iteration where the DEC based Richardson updates take over the Jacobi updates, and the pink dotted line represents the continued residual trajectory of the Jacobi iterations. A sharp fall in the relative residual can be observed at this point, which can be attributed to the physics consistent DEC operator contributing to the solution approximation updates. The gradual increasing trend thereafter can be attributed to the update nature of Richardson iterations resulting in incrementally mild over-corrections. The residual spikes in \ref{fig:poisson_res}(b) occur during DeepONet corrections of the HINTS algorithm. This is as expected, as the DeepONet is trained to map inputs to the solution function space, and not optimized through residual minimization. The Jacobi updates show a continuous decrease in the iteration residual norm due to the contractive nature of its update rule.


\subsection{Helmholtz Equation}
\noindent The 2D Helmholtz equation is written as:
\begin{equation}\label{helmholtz}
\begin{alignedat}{2}
(\alpha - \Delta)  u (x,y) & = f(x,y) \quad && \forall \;\; (x,y)\in \Omega, \\
u(x,y)        & = 0  \quad && \forall \;\; (x,y)\in \partial\Omega,
\end{alignedat}
\end{equation}
where $\alpha$ is the wave number. We use the same approach as that of the Poisson Equation for generating labeled data. The synthetic solution-forcing function pairs are generated using
\begin{equation}\label{fourier-helmholtz}
\begin{aligned}
    u(x,y) &= \frac{1}{N}\sum_{n=1}^Na_nsin(n\pi x).cos(n \pi y), \\
    f(x,y) &= \frac{1}{N}\sum_{n=1}^N(2a_n(n\pi)^2  \ + \ \alpha)sin(n\pi x)\cdot cos(n \pi y).
\end{aligned}
\end{equation}
We consider the same domain as the Poisson problem, $x \in [0,1], \ y \in [-0.5,0.5]$, and $\alpha \geq 2^2$. Our PDE representation for the algorithm is
\begin{equation}
    \mathcal{L}(u) \ = \ \alpha u \ - \ (\Delta_0u \ + \ \mathcal{CNN}(\Delta_0u)).
\end{equation}
The corresponding Jacobi iteration stencil with grid spacing $h$ reads
\vspace{0.3cm}
\begin{equation}
    \frac{1}{h^2}
\begin{bmatrix}
0 & -1 & 0 \\
-1 & 4 + \alpha h^2 & -1 \\
0 & -1 & 0
\end{bmatrix}.
\end{equation}

\begin{figure}[!h]
    \centering
    \includegraphics[width=1\textwidth]{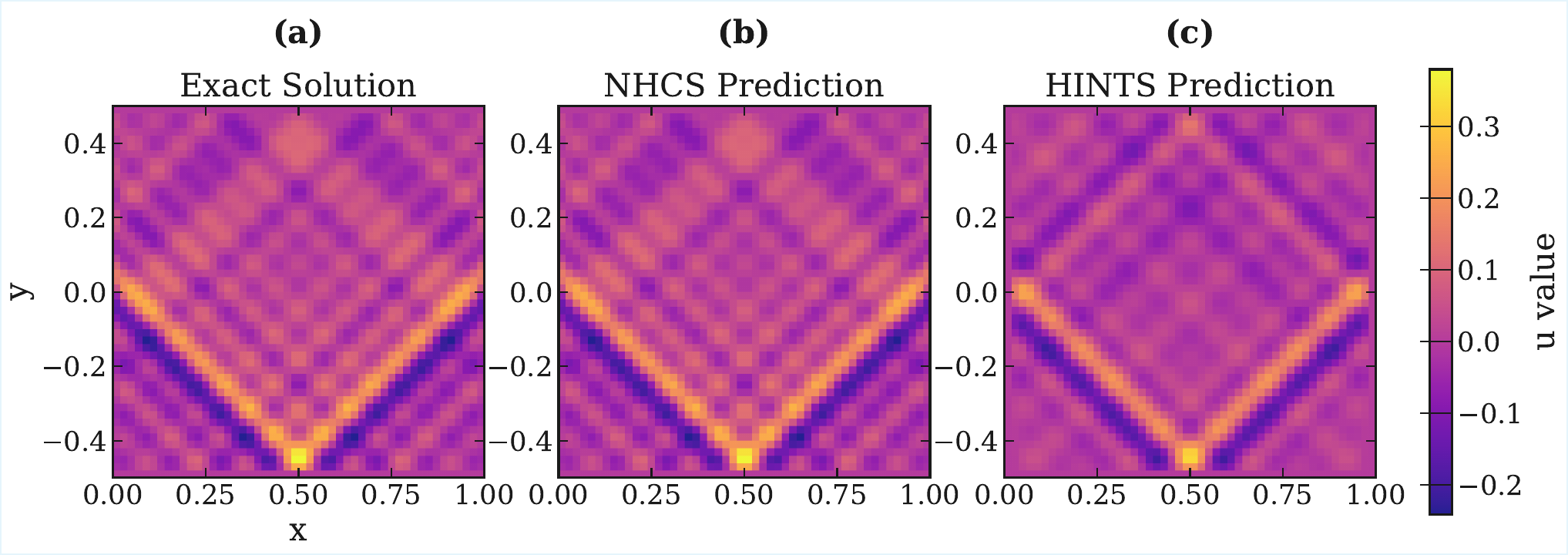}
    \caption{Helmholtz equation solution estimate comparison between (b) NHCS, and (c) HINTS with respect to the exact solution (a). Again, NHCS approximates the solution field more accurately than HINTS, which tends to only capture the dominant features of the solution field in high potential regions, while over-smoothing local high-frequency variations.}
    \label{fig:helm_solns}
\end{figure}
\begin{figure}[!h]
    \centering
    \includegraphics[width=1\textwidth]{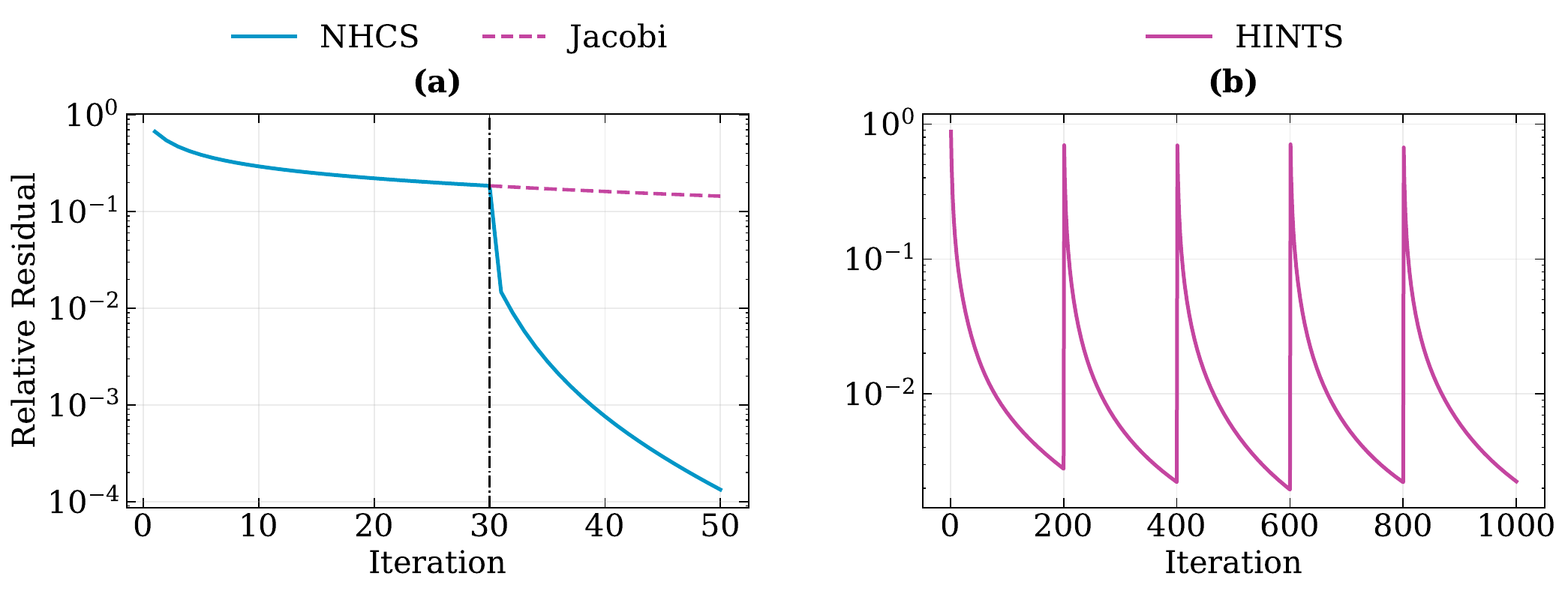}
    \caption{Helmholtz equation relative residual vs iteration count for NHCS (a) and HINTS (b). In (a) the blue line represents the residual norm of NHCS, the black dotted line represents the iteration where the DEC based Richardson updates take over the Jacobi updates, and the pink dotted line represents the continued residual trajectory of the Jacobi iterations. The sharp fall in the residual in (a) can be attributed to the physics consistent DEC operators contributing to the correction during the Richardson updates and the spikes in (b) occur during DeepONet corrections in HINTS.}
    \label{fig:helm_res}
\end{figure}

Figure~\ref{fig:train}(c, d), and Table~\ref{tab:training_comparison} summarizes the Phase 1 training performance across different mesh sizes. Similar to the Poisson equation, the time taken for training increases with the discretization density of the mesh. Figure~\ref{fig:train2}(c, d), and Table~\ref{tab:training_comparison2} summarize the Phase 2 training performance, and a $\mathcal{N}:\mathcal{M}$ = $30:20$ configuration is used. Figure~\ref{fig:runtime}(b) illustrates the computational time required to evaluate solutions as a function of grid size. Again, similar to the Poisson equation, a gradual increase in runtime is observed for HINTS as the grid resolution increases from $30 \times 30$ to $64 \times 64$, while the NHCS runtime remains nearly constant over the same range of grid sizes, indicating weak sensitivity to grid refinement. Figure~\ref{fig:comparison}(b) compares the evaluation performance between NHCS and HINTS for the Helmholtz equation. The error generally decreases with mesh refinement, except for the $35\times35$ case, reaching a minimum of approximately $6\times10^{-4}$ at $64\times64$. Notably, even the $35\times35$ model achieves a low error of approximately $1\times10^{-3}$. Conversely, HINTS displays a steady rise in error up to $50\times50$, followed by a sharp increase to nearly $0.9$ at $64\times64$. The non-monotonic behavior of NHCS can be attributed to the presence of the
wavenumber term alters the spectral structure of the operator, causing the relative contribution of oscillatory modes to vary with resolution. Therefore, certain mesh sizes align more favorably with the learned operator representation, producing lower errors at intermediate resolutions. Figure~\ref{fig:helm_solns} visualizes and compares the solution approximation capabilities of HINTS and NHCS for a randomly generated sample. Similar to the results seen for the Poisson equation, NHCS is able to approximate the field representation accurately. In contrast, HINTS captures only certain dominant features, notably the high potential diagonal region, while over-smoothing other regions, resulting in a visibly less refined and inconsistent approximation. Figure \ref{fig:helm_res}(a) and (b) capture the change in relative residual with respect to iteration count for NHCS and HINTS respectively. Similar to the Poisson equation, a sharp increase can be observed after the Richardson updates take over in NHCS. Unlike the Poisson equation however, it progressively decreases thereafter. Again, like the Poisson equation, HINTS shows residual spikes when DeepONet corrections take place.

\subsection{Linear Elasticity Equation}

We consider a square plate of size $1\,\text{m} \times 1\,\text{m}$ with a $0.2\,\text{m} \times 0.2\,\text{m}$ centered, square cutout. The bottom edge of the plate is fixed (zero displacement), and a uniform vertical displacement $\Delta_y$ is applied on the top edge, while the left and right edges are traction-free. The body force, $f$, is set to zero. The 2D linear elasticity problem is governed by the PDE, written as:
\begin{equation}\label{elasticity}
\begin{alignedat}{3}
&\left\{
\begin{aligned}
&\nabla \cdot \sigma(x,y) + f(x,y) = 0 \\
&\sigma(x,y) = C : \varepsilon(x,y) \\
&\varepsilon = \tfrac{1}{2}\left(\nabla u + (\nabla u)^\top\right)
\end{aligned}
\right.
&& \text{in } \Omega, \\[6pt]
\end{alignedat}
\end{equation}
such that $u = 0$ on $\partial\Omega_b$, $u_y = \Delta_y$ on $\partial\Omega_t$, and $\sigma \cdot n = 0$ on  
$\partial\Omega_l \cup \partial\Omega_r \cup \partial\Omega_h$,
where ${u}(x,y) = [u_x, u_y]^\top$ is the displacement field, $\boldsymbol{\varepsilon}$ is the strain tensor, $\Delta_y$ is the applied displacement, and $\boldsymbol{\sigma}$ is the Cauchy stress tensor. $\partial \Omega_t$, $\partial \Omega_b$, $\partial \Omega_l$, $\partial \Omega_r$, and $\partial \Omega_h$ denote the top, bottom, left, right and hole boundaries of the plate, respectively. The fourth-order elasticity tensor $\mathcal{C}$ is defined by the material properties, Young’s modulus $E$ and Poisson’s ratio $\nu$ as:
\begin{equation}
\mathcal{C} = \frac{E}{1-\nu^2}
\begin{bmatrix}
1 & \nu & 0 \\
\nu & 1 & 0 \\
0 & 0 & \tfrac{1-\nu}{2}
\end{bmatrix}.
\end{equation} 
For generating data, we uniformly sample the Young's Modulus within the bounds $[7\text{MPa}, 10\text{MPa}]$, the Poisson's ratio from $[0.15, 0.35]$, and $\Delta_y$ from $[0.01, 0.25]$, and target solutions are generated using the PDE solver of Matlab. The 1-chain-wise DDEC representation of the stress tensor is written as:
\begin{equation}
    {\varepsilon}_k(u) 
= \frac{1}{2}
\begin{bmatrix}
2\,(\text{GRAD}\,u_x)_k & (\text{GRAD}\,u_y)_k \\[6pt]
(\text{GRAD}\,u_y)_k & 2\,(\text{GRAD}\,u_y)_k
\end{bmatrix},
\end{equation}
and the PDE representation for the algorithm is
\begin{equation}\label{ddec_elas}
\begin{alignedat}{1}
\lambda &= \frac{E\nu}{(1+\nu)(1-2\nu)}, \\
\mu &= \frac{E}{2(1+\nu)},\\
\mathcal{K}u &= \text{GRAD}^* \mathcal{C}:\varepsilon(u)\\
            &= \text{GRAD}^*(\lambda\text{tr}(\varepsilon(u))\mathbf{I} \ + \ 2\mu\varepsilon(u ), \\
\mathcal{L}(u) &= \mathcal{K}u \ + \ \mathcal{CNN}(\mathcal{K}u).
\end{alignedat}
\end{equation}

\begin{table}[!ht]
    \centering
    \begin{tabularx}{0.8\textwidth}{c *{2}{>{\centering\arraybackslash}X}}
        \toprule
        Grid Size & RMSE & Time (s) \\
        \midrule
        $30 \times 30$ & $9.927 \times 10^{-5}$ & $58.77$ \\
        $35 \times 35$ & $7.446 \times 10^{-5}$ & $68.62$ \\
        $40 \times 40$ & $5.039 \times 10^{-5}$ & $72.93$ \\
        $45 \times 45$ & $7.726 \times 10^{-5}$ & $107.65$ \\
        $50 \times 50$ & $4.266 \times 10^{-5}$ & $146.21$ \\
        $60 \times 60$ & $9.000 \times 10^{-5}$ & $271.21$ \\
        \bottomrule
    \end{tabularx}
    \caption{Phase 1 training performance (RMSE and training time) for the Elasticity problem (plane stress with central hole) across different grid sizes over 500 epochs.}
    \label{tab:elas_training}
\end{table}

\begin{table}[!ht]
    \centering
    \begin{tabularx}{\textwidth}{c *{2}{>{\centering\arraybackslash}X}}
        \toprule
        Grid Size & Relative MSE & Time (s) \\
        \midrule
        $30 \times 30$ & $2.597 \times 10^{-3}$ & $262.52$  \\
        $35 \times 35$ & $3.012 \times 10^{-3}$ & $447.77$ \\
        $40 \times 40$ & $2.473 \times 10^{-3}$ & $537.76$ \\
        $45 \times 45$ & $3.315 \times 10^{-3}$ & $694.60$ \\
        $50 \times 50$ & $2.248 \times 10^{-3}$ & $820.62$ \\
        $60 \times 60$ & $2.008 \times 10^{-3}$ & $1431.45$ \\
        \bottomrule
    \end{tabularx}
    \caption{Phase 2 training performance (Relative MSE and training time) across different grid sizes over 1000 epochs.}
    \label{tab:elas_training2}
\end{table}

Figure~\ref{fig:train}(e, f), and Table~\ref{tab:elas_training} summarizes the Phase 1 training performance across different mesh sizes. Similar to the Poisson and Helmholtz equations, the time taken for training increases with the discretization density of the mesh. Figure~\ref{fig:train2}(e, f), and Table~\ref{tab:elas_training2} summarize the Phase 2 training performance where a modified $1:7$ configuration is employed. Instead of running a single Jacobi update, we begin by considering the square domain with no hole and approximate the solution by applying a uniaxial extension in the y direction, and a lateral contraction determined by Poisson’s ratio in the x direction. If $\Delta_y$ is the top layer y-axis displacement and $\nu$ is the Poisson's ratio, then the initial approximation is computed as
\begin{equation}
\begin{aligned}
u_x(x,y) &= -\,\nu\,\frac{\Delta}{L}\,x, \\
u_y(x,y) &= \phantom{-}\frac{\Delta}{L}\,y,
\end{aligned}
\end{equation}
where $u_x \text{and} u_y$ are the x and y direction displacements respectively. This initial approximation is then passed into the Richardson update loop. Figure~\ref{fig:train2}(f) illustrates a generally decreasing trend of final training error with increasing model size for the elasticity problem. Although geometric discontinuities in the domain can cause local error accumulation, finer meshes allow the model to capture these variations more accurately and resolve the resulting approximation errors. Furthermore, the use of a semi-analytic initial approximation results in a more consistent convergence behavior across all training samples.

\begin{figure}[!h]
    \centering
    \includegraphics[width=1\textwidth]{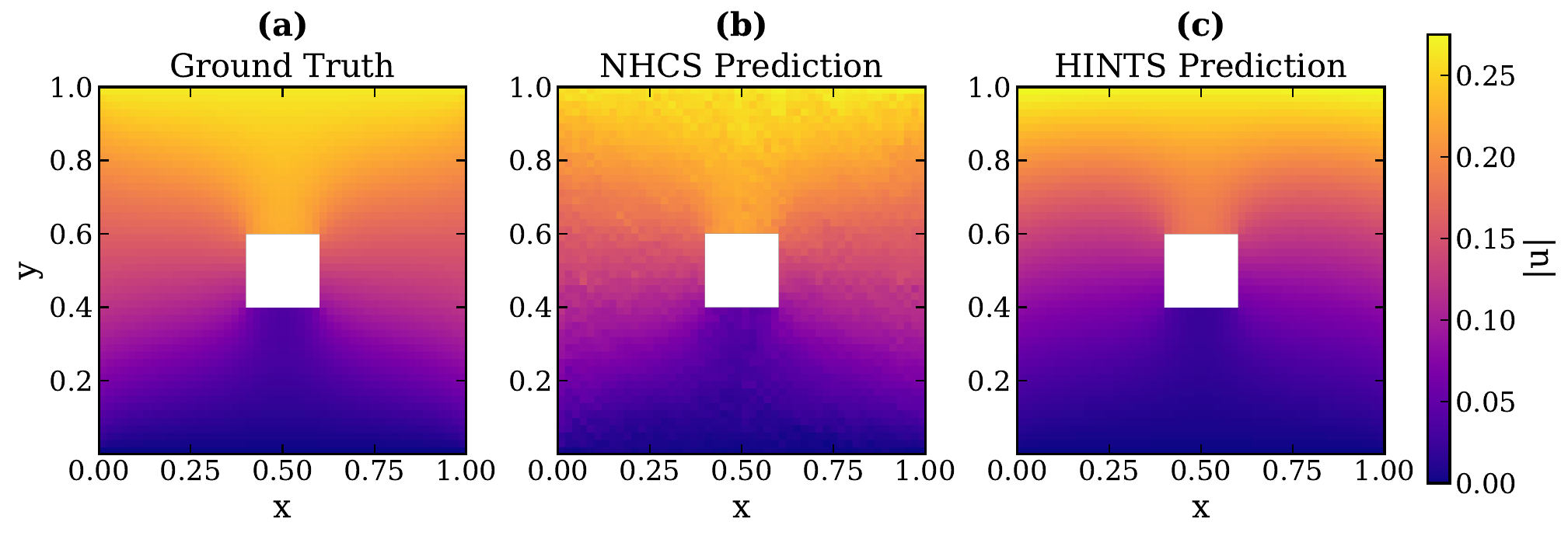}
    \caption{Elasticity problem solution estimate comparison between (b) NHCS, and (c) HINTS with respect to the exact solution (a). Coarse localized errors are clearly visible in the NHCS approximation. Regardless, it accurately approximates the solution field despite the geometric discontinuity. The HINTS approximation is significantly smoother, however the error increases closer to the hole boundary. The NHCS prediction can be further refined with a few additional Jacobi iterations to smooth out coarser errors.}
    \label{fig:elas_solns}
\end{figure}

\begin{figure}[!h]
    \centering
    \includegraphics[width=1\textwidth]{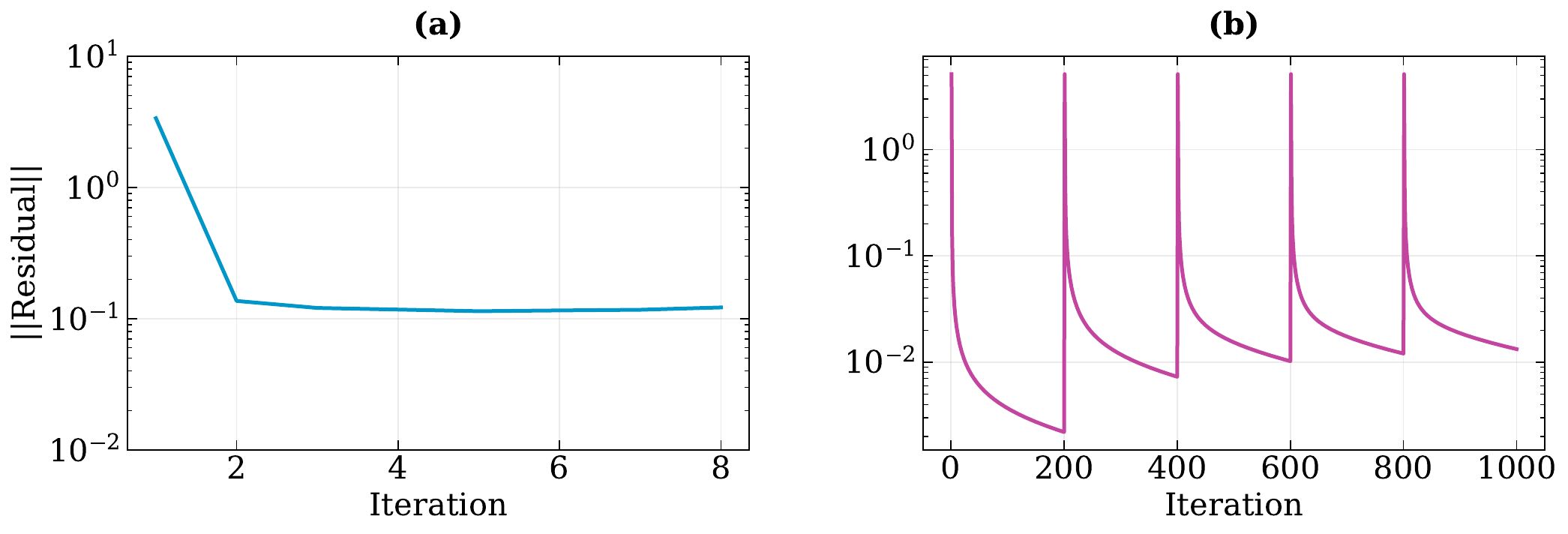}
    \caption{Elasticity problem iteration residual norm vs iteration count for NHCS (a) and HINTS (b). The iteration residual evolution in (a) is significantly more stable as compared to the Helmholtz and Poisson equation, due to the low frequency nature of the solution, resulting in stabler updates. The spikes in (b) are again due to DeepONet corrections in the solution space.}
    \label{fig:elas_res}
\end{figure}

Figure~\ref{fig:runtime}(c) illustrates the computational time required to evaluate solutions as a function of grid size. Here the performance difference becomes particularly pronounced where the HINTS runtime increases from approximately $8$~s at a $30 \times 30$ resolution to over $30$~s at $60 \times 60$. Although NHCS exhibits a mild growth trend, its runtime remains below $0.06$~s across all grid sizes, resulting in speedups of several orders of magnitude compared to HINTS. A zoomed in log-scale plot for the NHCS runtime has also been provided for more clarity. Figure~\ref{fig:comparison}(c) compares the evaluation performance between NHCS and HINTS for the 2D Elasticity equation problem. As expected from the Phase 2 training plots, a generally decreasing trend in error for NHCS can be observed. In contrast, HINTS sees an increasing relative MSE under mesh refinement, consistent with our previous observations for the Poisson and Helmholtz problems. With the sole exception of the $45 \times 45$ case, NHCS maintains an error on the order of $10^{-3}$, observing a minimum error of under $4 \times 10^{-3}$ on a $60 \times 60$ mesh. HINTS does exhibit higher accuracy than NHCS for the $30 \times 30$ and $35 \times 35$ meshes; however, as the grid resolution increases, its error rises sharply, reaching nearly $0.08$ for the $60 \times 60$ case. The governing equations form a coupled PDE system, and mesh refinement alters how this coupling is discretized and corrected by the DEC operator during the Richardson updates, thus resulting in the non-monotonic behavior of NHCS. Figure~\ref{fig:elas_solns} illustrates the performance of the two methods by visualizing the predicted solutions for a randomly generated sample. NHCS is able to capture the geometric inconsistencies, albeit with some localized errors. It accurately approximates the solution, although some regions remain less smooth. HINTS approximates a smoother displacement field, yet  it is visibly less accurate with errors accumulating closer to the hole boundary. The DeepONet component also seems to overshoot the top boundary displacement approximation in the final iteration. This demonstrates the adaptibility of NHCS in problems with geometrical inconsistencies. Since the forcing term is zero we plotted the residual norm trajectory instead of the relative residual. Figure \ref{fig:elas_res}(a) and (b) visualize the iteration residual norm trajectory as compared to the iteration count for NHCS and HINTS respectively. The iteration residual evolution in (a) is significantly more stable as compared to the Helmholtz and Poisson equation, due to the low frequency nature of the solution, resulting in stabler updates. The spikes in (b) are again due to DeepONet corrections in the solution space, with the Jacobi iterations showing a monotonically decreasing trend due to the contractive nature of its update. This experiment demonstrates that NHCS naturally extends beyond scalar elliptic problems to vector-valued, coupled PDEs with complex geometry, without architectural changes, as the operator accounts for the structure of the domain.

\section{Conclusions and Discussions}\label{sec:conclusion}
We have introduced the Neural Hodge Corrective Solver (NHCS), a hybrid neural-iterative framework for solving PDEs that leverages discrete exterior calculus (DEC) operators and data-driven metric learning. The PDE operator is constructed from DEC co-boundary and codifferential operators, while a convolutional correction term captures local and global variations, providing multiscale adaptivity. NHCS integrates classical iterative solvers with learned corrections in a principled manner, using Jacobi iterations to efficiently reduce high-frequency errors and Neural Hodge corrective iterations to resolve low-frequency and multiscale features. The staggered two-phase training ensures stable and interpretable learning of the DEC operators, while the evaluation procedure avoids expensive Jacobian computations and reduces memory overhead. Through numerical experiments on Poisson, Helmholtz, and linear elasticity problems, we demonstrated that NHCS preserves both topological and metric structure, accurately captures high- and low-frequency solution components, and maintains sharp solution features even in the presence of geometric discontinuities. Compared to HINTS, NHCS exhibits faster convergence, improved solution fidelity, and better multiscale resolution, with residual trajectories indicating more effective iterative updates within fewer iterations.

The framework is lightweight, highly adaptive, and offers a unified approach to integrating classical and learned components, making it broadly applicable to PDEs with complex solution structures. By combining structure-preserving operators with a single convolutional correction term, NHCS achieves a balance between computational efficiency and numerical accuracy, avoiding the over-smoothing and high computational cost observed in previous hybrid methods. The proposed approach demonstrates that principled incorporation of geometric and topological structure into learned solvers can enhance interpretability, convergence, and robustness, pointing to a promising direction for hybrid neural-numerical PDE solvers.

While NHCS shows strong performance on regular two-dimensional domains and problems with moderate geometric complexity, several extensions remain as future directions. Scaling to larger meshes or unstructured grids is currently limited by the computational cost of constructing and applying the coboundary operator, which scales as \(\mathcal{O}(n^4)\) with the number of elements. Domain decomposition and parallelization, or encoding strategies that map irregular meshes onto coarser representations, could address these challenges. Additionally, extending NHCS to time-dependent or nonlinear PDEs, potentially through stable hybrid time integrators, is a promising avenue, with recent advances in discrete exterior calculus on spacetime meshes suggesting feasibility. Finally, leveraging learned adaptive preconditioners within the hybrid framework may further accelerate convergence, drawing on the success of DEC-based methods in manifold learning and data analysis.

\section*{Acknowledgements}
S. Chakraborty acknowledges the financial support received from the Ministry of Port and Shipping via letter no. ST14011/74/MT (356529), and seed grant received from IIT Delhi. 

\section*{Competing interests} 
The authors declare that they have no conflict of interest.

\newpage
\appendix
\section{Architectural Specifications}\label{sec:app1}

\begin{table}[h!]
\centering
\label{tab:hints_branch}
\renewcommand{\arraystretch}{1.25}
\begin{tabular}{p{2.8cm} p{3.2cm} p{4.5cm} p{2.8cm} p{2.8cm}}
\hline
\multicolumn{5}{c}{\textbf{Branch Net Specifications}} \\
\hline
\textbf{Problem} 
& \textbf{Input Channels} 
& \textbf{Convolutional Channels} 
& \textbf{Kernel / Stride} 
& \textbf{Linear Layers} \\
\hline

Poisson 
& $1$ 
& $[1, 40, 60, 100]$ 
& $3 / 2$ 
& $[80, 80, 80]$ \\

Helmholtz 
& $2$ 
& $[2, 40, 60, 100]$ 
& $3 / 2$ 
& $[80, 80, 80]$ \\

Elasticity 
& $5$ 
& $[5, 40, 60, 100]$ 
& $3 / 2$ 
& $[80, 80, 80]$ \\

\hline
\end{tabular}
\caption{Branch Net architecture specifications for the DeepONet used in the HINTS framework. The output of the convolutional layers is projected to a final output of size 80 through a set of three fully connected layers with ReLU activation functions between each layer.}
\end{table}

\begin{table}[h!]
\centering
\label{tab:hints_trunk}
\renewcommand{\arraystretch}{1.25}
\begin{tabular}{p{2.8cm} p{3.5cm} p{3.5cm} p{3cm} p{3cm}}
\hline
\multicolumn{5}{c}{\textbf{Trunk Net Specifications and DeepONet Training Hyperparameters}} \\
\hline
\textbf{Problem} 
& \textbf{Input Dimension} 
& \textbf{Linear Layers} 
& \textbf{Activation} 
& \textbf{Training Epochs / LR} \\
\hline

Poisson 
& $2$
& $[80, 80, 80]$ 
& Tanh 
& $10{,}000$ / $1 \times 10^{-3}$ \\

Helmholtz 
& $2$
& $[80, 80, 80]$ 
& Tanh 
& $10{,}000$ / $1 \times 10^{-3}$ \\

Elasticity 
& $2$
& $[80, 80, 80]$ 
& Tanh 
& $20{,}000$ / $5 \times 10^{-3}$ \\

\hline
\end{tabular}
\caption{Trunk Net architecture specifications and DeepONet training parameters used in HINTS.}
\end{table}

\end{document}